# Towards the issue of the origin of Fermi surface, pseudogaps and Fermi arcs in cuprate HTSCs

Kirill Mitsen[*] & Olga Ivanenko
Lebedev Physical Institute, Leninsky pr., 53 Moscow, 119991, Russia

**Abstract.** Earlier we have proposed a new approach to the analysis of phase diagrams for cuprates and pnictides and have shown that the positions of superconducting domes can be predicted proceeding from only the crystal structure of a compound. This approach uses the concept of the self-localization of doped carriers due to their formation of trion complexes that represent a bound state of the doped carrier and charge transfer excitons emerging under its influence. Here, as exemplified by cuprates, we show that the use of the proposed approach to the analysis of the transformation of an electronic structure with doping enables an explanation to a range of their anomalies: Fermi arcs, "large" and "small" pseudogaps etc. The basic conclusion is that the role of the Fermi surface in cuprates is played by an isoenergetic contour that emerges at the sectioning of the surface of a band dispersion by a dispersionless "biexciton" pair level. This level additionally plays the role of an acceptor to lead to the emergence of hole carriers on the isoenergetic contour and to a jump of the chemical potential. Based on the conducted consideration, we propose a possible mechanism of superconducting pairing genetically inherent in such a system.

## Introduction

One of the most striking features of HTSCs is the emergence of the so-called pseudogap in the excitation spectrum, i.e., a sharp decrease in the density of states of spin and charge excitations at temperature $T^*$, which can greatly exceed $T_c$.

Experiments have provided important information about mysterious features of the pseudogap phase and interrelations between the pseudogap, superconducting gap and Fermi arcs. (i) There is a $d$-wave superconducting gap on the 2D Fermi surface of optimally doped HTSC cuprates [1]; (ii) The transition from the superconducting to normal state is accompanied by the closing of the superconducting gap across parts of the Fermi surface and the formation of Fermi arcs centered at the nodes. Meanwhile, the gap in the vicinity of antinode directions (pseudogap) closes at some temperature $T = T^*$ well above $T_c$ [2,3]; (iii) As the doping level is reduced, the pseudogap increases and, for $T<T_c$, the gap at the Fermi surface deviates from the simple $d$-wave behavior [4]. In deeply underdoped samples, Cooper pairs form islands in $k$-space around the nodal regions [5]; (iv) Above $T_c$, underdoped samples exhibit a number of anomalies such as a giant Nernst effect and anomalous diamagnetism [6,7].

Various viewpoints exist regarding the nature of the pseudogap and pseudogap anomalies. In the opinion of most investigators, the observed anomalies suggest the existence of strong superconducting fluctuations in the pseudogap phase and, thus, the pseudogap is as if a precursor of superconductivity. Another point of view is that the pseudogap and the superconducting gap in HTSCs are manifestations of two competing order parameters, because the effects of various factors, such as, e.g., doping and disordering, on these gaps are differently directed [8].

In this work we would like to show that the observed pattern of the formation of Fermi arcs and a pseudogap in cuprates can be understood within the framework of our earlier proposed model of heterovalent doping of cuprate and pnictide HTSCs.

The model has enabled predicting the positions of superconducting domes on the phase diagrams of both cuprates and pnictides proceeding only from the knowledge of their crystal structure and the position of the dopant in it [9]. According to the model as applied to cuprates, introduction of a dopant ion into the lattice of the parent insulator is accompanied with the emergence of an additional (doped) carrier (electron or hole) in the basal plane or near it. The carrier self-localizes in the nearest vicinity of the dopant due to the deformation of the local electronic structure, which leads to the formation of a trion complex comprising the doped carrier and charge transfer (CT) excitons generating in its vicinity. These CT excitons are generated in $CuO_4$ plaquettes in the $CuO_2$ plane at the boundary of the doped carrier localization region. Such a $CuO_4$ plaquette, where a CT exciton resonantly interacting with band states can be formed, we called a CT plaquette.

---

[*] Correspondence to [mitsen@lebedev.ru]

As the doping increases, these trion complexes get ordered due to Coulomb interaction between them. Accordingly, the arrangement of CT plaquettes in the basal plane of the crystal will be determined by its structure and the type of dopant so that the dopant concentration range corresponding to the existence region of the percolation cluster of CT plaquettes (the CT phase) can be readily determined for each particular compound. It has been shown [9] that for HTSC cuprates (as for HTSC pnictides) whose phase diagrams can be considered to be well established, these concentration ranges coincide exactly with the positions of superconducting domes on the experimental phase diagrams, i.e., namely the CT phase is an HTSC phase. This result suggests the existence of a common (both for cuprates and pnictides) and rather rough mechanism that acts irrespective of fine details of the band structure and provides for the superconducting pairing in these materials.

In this work we show that such a coincidence between the experimental and calculated superconductivity intervals on the HTSC phase diagrams is, apparently, not accidental. We demonstrate that the use of the proposed approach for the analysis of the electronic structure of cuprates enables a qualitative explanation to a whole range of their anomalous properties. For this, we first consider the transformation mechanism of their electronic structure from the state of an undoped CT insulator to an optimally doped superconductor. Further, based on this consideration, we propose a mechanism for the formation of the Fermi surface in cuprates (including the formation of "large" and "small" pseudogaps and Fermi arcs) and compare the anticipated results with the experiments. Simultaneously, we consider a mechanism for the generation of free carriers in cuprates at a heterovalent doping and propose a superconducting pairing mechanism genetically inherent in such a system.

## 1. The electronic structure of undoped cuprates. A remnant Fermi surface

The $Cu^{2+}$ ion in undoped cuprates is known to be in a $3d^9$ configuration, i.e., to have one hole in the d shell. Due to the crystal field of the oxygen octahedra and their extendedness in the direction of the *z* axis the 3d level splits up, as the result of which one unpaired electron proves on the upper $Cu3d_{x2-y2}$ level. As the energy levels of Cu3d and O2p are close, the $Cu3d_{x2-y2}$ level is strongly hybridized with four $O2p_{x,y}$ orbitals directed to the copper atom. If the tight binding approximation for O2p and $Cu3d_{x2-y2}$ orbitals (with five electrons per $CuO_2$ cell) is used for the analysis of the electronic structure, we will have three orbitals per unit cell and, therefore, three bands in a crystal. The lower band is bonding; the middle one, nonbonding; and the upper band is antibonding. As we have five electrons per cell, the Fermi level should be in the middle of the upper antibonding band. That is to say, the model predicts a metal state with the Fermi level in a half-occupied band, which is not consistent with the experiment as cuprates in an undoped state are known to be antiferromagnetic insulators. It is commonly accepted that the inability of band theory to correctly predict the properties of this compound is due to the neglect of electron correlation energy in calculations. These correlations lead to splitting the antibonding band into the upper (UHB) and lower (LHB) Hubbard subbands and the emergence of an energy gap on the Fermi level between the O2p valence band and the upper Hubbard subband formed mainly by copper orbitals [10-12]. The fact that this gap is much smaller than Hubbard energy, $U$, makes the cuprates charge-transfer insulators [13,14].

The experimentally observed [15] electron spectrum $E(\mathbf{k})$ of undoped cuprates is shown in Fig. 1a. Its distinctive features are the position of the maximum of dispersion in the centre of the Brillouin zone at point $\mathbf{k} = (\pi/2;\pi/2)$ and a sharp drop in occupation numbers $n(\mathbf{k})$ at some (not isoenergetic) contour (the white line in Fig. 1a), the projection of which onto the Brillouin zone reproduces in general outline the Fermi surface of an optimally doped compound (i.e., the "remnant" Fermi surface).

To describe the further transformation of the electronic structure with doping, it is required first and foremost to understand the mechanism that leads to such a specific behaviour of $n(\mathbf{k})$. Our consideration will be based on the assumption that the motion of an electron with a large momentum $\mathbf{k}$ (close to the "remnant" Fermi surface) in the $CuO_2$ plane can be considered as a motion in a quasi one-dimensional potential.

The basis for the assumption of such a character of electron motion is that the $2p\pi$ band, which arises from the hybridization of the $O-2p_{x,y}$ orbitals perpendicular to the $O-2p$ ligand orbitals, has almost no overlap with the $2p\sigma$ band which has the same symmetry of the $Cu3d_{x2-y2}$ states [16,17]. Specifically, the amplitudes of the projections of the wave functions on the $2p\sigma$ and $2p\pi$ orbitals are concentrated around momentum lines $(0,0)-(0,\pi)$ and $(\pi/2;\pi/2)-(\pi;\pi)$, respectively.

So, the propagation of an electron (Fig. 1b) with momentum $\mathbf{k}_1$ in the node direction goes through $2p\pi$ orbitals along diagonal oxygen chains, and that of an electron with momentum $\mathbf{k}_2$ in the antinode directions goes through $2p\sigma$ orbitals by tunnelling in the direction of Cu–O bonds. The propagation of an electron with arbitrary momentum $\mathbf{k}$ in the $CuO_2$ plane can be presented as the superposition of motions along the node and antinode directions, i.e. alternately through $2p\pi$ and $2p\sigma$ orbitals, respectively (the broken line in Fig. 1b).

This motion can be considered as the motion in a quasi-one-dimensional periodic potential of the Kronig–Penney type, with various amplitudes of barriers, where the widths of boxes and the averaged amplitudes of barriers depend on the direction of vector $\mathbf{k}$. Herewith, the electronic dispersion $E(\mathbf{k})$ in the Brillouin zone can be presented as an integration of dispersions $E(k)$ in 1D bands for various directions of $\mathbf{k}$. Correspondingly, the value of dispersion $E(k)$ in the direction of $\mathbf{k}_1(\mathbf{k}_2)$ will be determined by the overlap integral $t$ ($t'$) between $2p\pi$ ($2p\sigma$) orbitals of the nearest oxygen ions in the node (antinode) direction. As $t > t'$, at a change of $\mathbf{k}$ from $\mathbf{k}_1$ to $\mathbf{k}_2$ the dispersion will decrease in accordance with the increase of the frequency of tunneling of an electron through the Cu sites (Fig. 1b). Due to the absence of unoccupied states on oxygen ions (which is equivalent to a strong Hubbard repulsion for an additional electron), these 1D bands can be considered as completely filled.

Cu–O chains (in the directions of the *x* and *y* axes) and oxygen chains (in the diagonal directions) serve as periodic barriers for the motion of an electron in the $CuO_2$ plane. In the former case, an electron with momentum $\mathbf{k}$ will experience a Bragg reflection at $|\mathbf{k}| = \pi/d_1 = \pi \cos(\pi/4 - |\alpha|)/a$, where $d_1$ is the distance between Cu–O chains along $\mathbf{k}$ direction, $a$ is the lattice constant of the $CuO_2$ plane and $\alpha$ is the angle between vector $\mathbf{k}$ and the direction of the O–O diagonals. In the latter case, reflections will take place at $|\mathbf{k}| = \pi/d_2 = \pi \cos(\alpha)/a\sqrt{2}$, where $d_2$ is the period of modulating potential of O–O chains along $\mathbf{k}$.

The values of Bragg band gaps in points $|\mathbf{k}| = \pi \cos(\pi/4 - |\alpha|)/a$ (reflections from Cu–O chains), as well as the values of $E(\mathbf{k})$, will depend on the average height of the barriers, i.e., on the frequency of tunneling of an electron through the Cu sites. The maximum Bragg gap and the minimum dispersion will be observed in the direction of Cu–O bonds ($\alpha = \pm \pi/4$), where the transfer of an electron from cell to cell is possible only by tunnelling (the overlap integral, $t' < t$). As $|\alpha|$ decreases, the value of dispersion will rise monotonically in agreement with the decrease of the frequency of tunnel transitions through Cu sites. At $\alpha = 0$, the Bragg reflections from Cu–O and O–O chains will take place for the same value of $|\mathbf{k}| = \pi/a\sqrt{2}$ at point $(\pi/2;\pi/2)$ that corresponds to the maximum of $E(\mathbf{k})$.

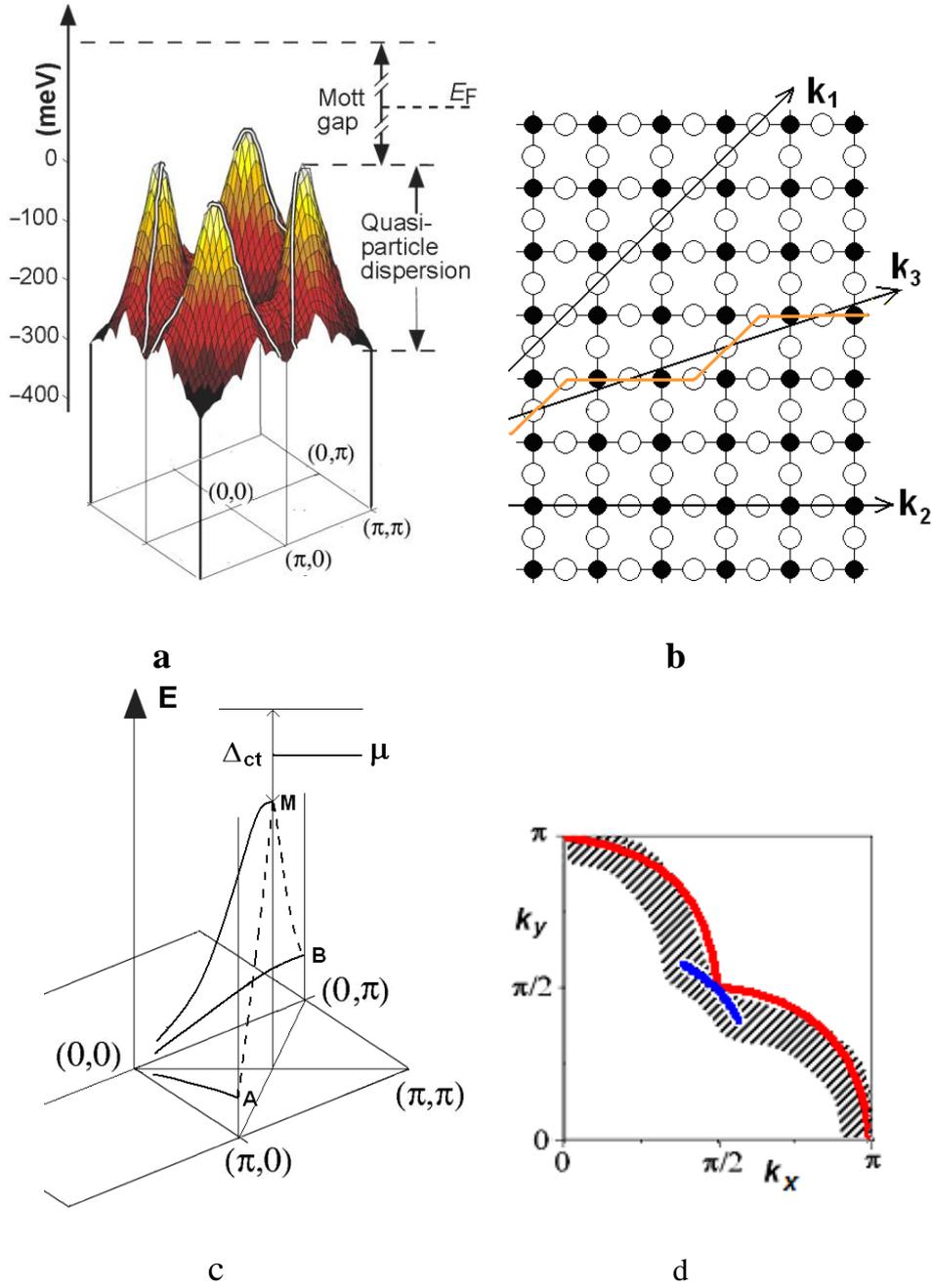

**Figure 1.** (**a**) Dispersion $E(\mathbf{k})$ measured by ARPES [15] for $Ca_2CuO_2Cl_2$. The white line in the figure designates a contour on which the occupation numbers $n(\mathbf{k})$ demonstrate a sharp decrease. The dispersion is maximum in the diagonal direction with the maximum $E(\mathbf{k})$ in point $(\pi/2,\pi/2)$ and is minimum in the direction of CuO bonds. (**b**) Propagation of an electron with arbitrary momentum $\mathbf{k}$ in the $CuO_2$ plane can be conditionally considered as the motion of an electron wave in one-dimensional potential, including the motion along diagonal oxygen chains and displacements, by means of tunnelling, from one chain to another (the brown broken line). (**c**) 1D dispersion curves for different directions of $\mathbf{k}$ expected from the proposed model. (**d**) The lines of a sharp decrease of the occupation numbers $n(\mathbf{k})$ in the 2D Brillouin zone of undoped cuprates: the hatched band, the region of $\mathbf{k}$ values in intersecting which $n(\mathbf{k})$ sharply decays [15]; the red and blue curves, the expected locus of 1D Brillouin zone boundaries.

It can be shown that the frequency of transitions of an electron with momentum $\mathbf{k}$ through the Cu sites in 1D potential is proportional to $(\cos(k_x a) - \cos(k_y a))$ (see further Fig. 5a). That is to say, the value of the dispersion in the direction from $(\pi,0)$ to $(\pi/2,\pi/2)$ will be proportional to $(\cos(k_x a) - \cos(k_y a))$, i.e., will be of a $d$-wave character. This pattern of band dispersion $E(\mathbf{k})$ (Fig. 1c) corresponds to the one observed in the experiment (Fig. 1a) [15].

In our model, due to the absence of unoccupied states on oxygen ions (which is equivalent to a strong Hubbard repulsion for an additional electron), these 1D bands for various *k* directions can be considered as completely filled. Therefore, the occupation numbers $n(k)$ should experience a sharp drop at the boundaries of 1D Brillouin zones at $|k| = \pi \cos(\pi/4 - |\alpha|)/a$, where the boundaries of 1D Brillouin zones are determined by reflections from Cu–O chains (the red curves) and at $|k| = \pi \cos(\alpha)/a\sqrt{2}$ (in the vicinity of $\alpha = 0$, where the boundaries of 1D Brillouin zones are determined by reflections from O–O chains, the blue curve). Constructed in this way, the common contour conforming to the expected position of the regions of a sharp decrease in the occupation numbers is in good agreement with the experimentally measured contour (the hatched area in Fig. 1d).

## 2. Transformation of the cuprate electronic structure under doping. "Large" pseudogap

Thus, in the initial state of the parent insulator the band states formed by nonbonding oxygen orbitals are separated from the electron states $3d^{10}$ on copper by the interband gap $\Delta_{ct}$ (Fig. 1). To describe the transformation of the cuprate electronic structure under doping within the framework of the model [9], let us recall its key points. According to the model, insertion of a dopant ion into the lattice of the parent insulator is accompanied with the emergence of an additional carrier (electron or hole) in the basal plane or near it. The charge of this doped carrier, in accordance with the symmetry of the environment, is spread over the nearest ions, imparting them with a fractional charge $q^* \approx \pm|e|/4$ (Fig. 2). At a hole doping a positive charge is induced on four nearest ions of O (Fig. 2a); at an electron doping, a negative charge, on four nearest ions of Cu (Fig. 2b). Note that in the case of $YBa_2Cu_3O_{6+x}$, for charge $q^* \geq \pm|e|/4$ to emerge on apical oxygen ions, excess oxygen should occupy two successive positions in the chain (Fig. 2c).

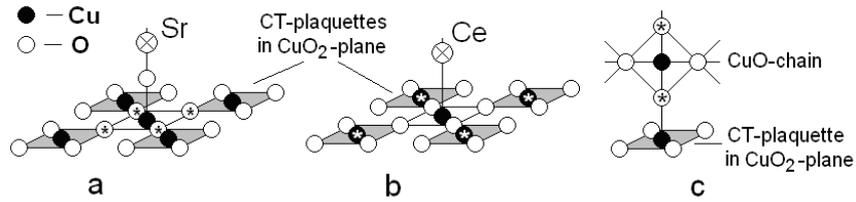

**Figure 2.** Formation of trion complexes, consisting of a localized doped carrier and four (LSCO, NCCO) or two (YBCO) plaquettes, in cuprates. The localization area of the doped carrier is restricted by the nearest ions: (**a**) oxygen ions, for the doped hole; (**b**) copper ions, for the doped electron. Ions onto which additional charge $q^* \approx \pm|e|/4$ sufficient to form a CT plaquette in the $CuO_2$ plane is induced are marked by stars. (**c**) in YBCO the doped carriers stay on apical oxygen ions.

In turn, each emerging induced charge $q^* \approx \pm|e|/4$ locally decreases the initial gap $\Delta_{ct}$ in its nearest vicinity down to a value $\Delta_{ct}^*$, which, according to estimates [18], enables an exciton transition with charge transfer (CT) between the Cu ion in the $CuO_2$ plane and neighbouring oxygen ions. This is possible when $\Delta_{ct}^* < E_{ex}$, where $E_{ex}$ is CT exciton binding energy (Fig. 3b). This hydrogen-like ionic complex ($CuO_4$) in the $CuO_2$ plane, which includes a Cu ion with surrounding O ions, where a CT exciton resonantly interacting with the non-bonding band states can arise, we called a **CT plaquette** (Fig. 2).

As each CT exciton in the vicinity of charge $q^* \approx \pm|e|/4$ is bound to it by Coulomb attraction, the doped carrier self-localizes due to the formation of a trion complex, which includes the carrier $\pm|e|$ proper and its surrounding CT plaquettes. These trion complexes get ordered in the basal plane due to Coulomb repulsion between them.

The arrangement of trion complexes in the basal plane of a crystal is determined by its structure and the type of dopant, so the dopant concentration range corresponding to the existence of a percolation cluster of CT plaquettes (the CT phase) can be readily determined for each particular compound. It has been shown [9] that for HTSCs, the phase diagrams of which can be considered to be well established, the indicated concentration ranges exactly coincide with the positions of superconducting domes on experimental phase diagrams. That is to say, it is the CT phase that is an HTSC phase. Note that in the cases when doped carriers are localized outside the $CuO_2$ plane (as, e.g., in YBCO, BiSrCaCuO etc.), it is possible to achieve such a level of doping when the CT phase occupies the entire $CuO_2$ plane. In this context, noteworthy is the work [19], where the authors, using *ab initio* quantum calculations at the level that leads to accurate band gaps, found that four-Cu-site plaquettes are created in the vicinity of dopants. They also calculated the threshold, at which the plaquettes percolate, so that the $Cud_{x2-y2}/Op\sigma$ orbitals inside the plaquettes now form a band of states along the percolating swath.

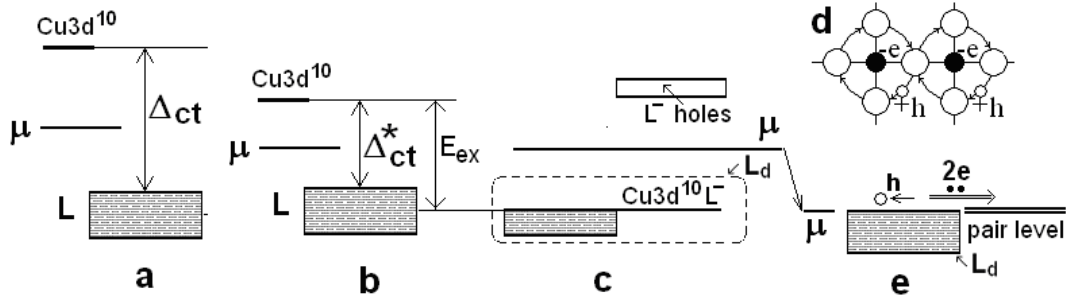

**Figure 3.** Transformation of the electronic structure of cuprates under doping. (**a**) Undoped phase (CT insulator). The states of electrons on copper $Cu3d^{10}$ are separated from the occupied nonbonding oxygen band by a gap $\Delta_{ct}$; (**b**) the doping reduces the gap $\Delta_{ct}$ down to $\Delta^*_{ct} < E_{ex}$, where $E_{ex}$ is CT exciton binding energy; (**c**) at $\Delta^*_{ct} < E_{ex}$, electrons from the upper part of L band transit into exciton-bound states $Cu3d^{10}L^-$, and the arising holes form an L$^-$ band. The states of electrons in the band and in the exciton-bound state $Cu3d^{10}L^-$ form a composite $L_d$ band, the states in which are a superposition of the band and exciton-bound states; (**d**) each pair of neighbouring CT plaquettes forms an HL centre at the pair level of which two electrons of $Cu3d^{10}L^-$ can form a bound state due to the possibility for holes to be attracted to both cations; (**e**) this HL centre can act as an acceptor, being ionized similar to the molecular ion $H_2^+$ and generating a free hole in L band. The emergence of hole carriers will be accompanied with a jump of chemical potential $\mu$ to the position of the pair level at the top of $L_d$ band.

Thus, the main result of doping is that the electronic structure of the $CuO_2$ plane is transformed from the state of the CT insulator (Fig. 3a) into the CT phase (Fig. 3b) in which the band states can resonantly interact with the exciton states. That is to say, at $\Delta^*_{ct} < E_{ex}$ (Fig. 3b) electrons occupying the states in the upper part of the oxygen L band will pass into exciton states $Cu3d^{10}L^-$, where $L^-$ denotes a hole on oxygen neighbouring with the copper ion (Fig. 3c). Correspondingly, the emerging oxygen holes will occupy states in L$^-$ band higher than the Fermi level. The formation of such a hole oxygen subband at the doping is confirmed by observing an additional absorption peak, 0.5 eV higher than the Fermi level, in the electron energy loss spectrum of $La_{2-x}Sr_xCuO_4$ [20]. As an issue important in below considerations, we note here that, owing to the dispersionless character of the exciton level of $Cu3d^{10}L^-$, we will have an isoenergetic contour in the cross section of L band. Electrons on this contour (band electrons) can pass into exciton $Cu3d^{10}L^-$ states and back, i.e., will be a superposition of the band and exciton states. Note that herewith the system still remains an insulator. To understand how such a system becomes a conductor and from where free carriers emerge, let us consider a composite electron band $L_d$ being the superposition of the electron states in L band and at the exciton level of $Cu3d^{10}L^-$ (Fig. 3c).

Let there be two CT plaquettes centred on the nearest cations of Cu (Fig. 3d). Such a pair of CT plaquettes represents a solid state analogue of the hydrogen molecule [18] and can be considered as a Heitler–London (HL) centre. On such a centre two electrons of $Cu3d^{10}L^-$ with opposite spins on the central ions of Cu can form a bound state (lying energetically lower than $Cu3d^{10}L^-$) due to the possibility of singlet holes (with necessity emerging in the CT phase on the surrounding oxygen ions); be spaced between Cu ions; and be attracted simultaneously to two electrons occurring on these ions. An additional decrease of energy, $\Delta E_{HL}$, in this case can be estimated from the ratio $\Delta E_{HL} \sim \Delta E_{H2}/\varepsilon_\infty^2 \approx 0.2$ eV, where $\Delta E_{H2} = 4.75$ is bond energy in the molecule of $H_2$, and $\varepsilon_\infty \approx 4.5$–5 (for cuprates).

Thus, in $L_d$ band two electrons on Cu cations will have a lower energy if they occupy neighbouring cations and two holes bound to them are on the surrounding O anions. These electrons occupy the common pair level of HL centres belonging to one percolation cluster (Fig. 3e). This **dispersionless** pair level of bound electrons can be observed in the ARPES as a flat band lying below the Fermi level.

At the same time, apart from the molecule of $H_2$, an analogue of which is the HL centre, the bound state of two protons and one electron, namely, the $H_2^+$ ion, is known to exist. In our case, it corresponds to an ionized HL centre, on which two electrons and one hole form a bound state. That is to say, the HL centre can act as an acceptor. As the result, free hole carriers will appear on the isoenergetic contour and the level of the chemical potential will go down to the top of $L_d$ band (Fig. 3e). Herewith, **the isoenergetic contour produced by sectioning the electronic structure by the dispersionless exciton level of $Cu3d^{10}L^-$ will play the role of a Fermi surface essentially not being such a surface.**

Thus, we are talking of the existence of two electronic subsystems: localized doped carriers and free holes [21]. The former form HL centres on pairs of adjacent Cu cations, the latter emerge at the occupation of these HL centres with electrons.

The concentration of free hole carriers will be determined from the condition of equality of the chemical potentials for electron pairs on the pair level and of electrons in L band. For small concentrations when the occupation of separate HL centres can be considered to be independent one of another, the change of the chemical potential for electron pairs due to transition of part of electrons from a band to HL centres can be presented as [22]:

$$\Delta\mu_p = -\frac{T}{2}ln[f(n)], \qquad (1)$$

where $n$ is the two-dimensional concentration of formed free holes per Cu ion, and $f(n)$ is some function of occupation of HL centres. Herewith, the change of the chemical potential for band electrons $\Delta\mu_e \approx -n/N(0)$, where $N(0)$ is the density of states on the Fermi level in a band. Equating the chemical potential changes of pairs and band electrons, we find in an explicit form the dependence of hole concentrations $n$ on temperature:

$$T \approx 2n/N(0)ln[f(n)]. \qquad (2)$$

Thus, in CT phase at low $T$, $n \propto T$. At high $T$, the concentration tends to a saturation, when the population of copper sites tends to unity (i.e., all HL centres are occupied). Comparison of the doping and temperature dependences of the measured Hall effect in $YBa_2Cu_3O_{6+x}$ [23] with predictions of the proposed model was made in [24].

In the case of optimal doping due to the interaction of electrons with HL centres, the distribution of free hole carriers proves nondegenerate in the sense that the chemical potential for holes $\mu = 0$ for all $T$, whereas the degeneration condition is the requirement that $\mu > 0$. Therefore, for holes all occupation numbers are less than unity and are distributed in the zone $\sim \Gamma$ ($\Gamma$, the value of two-particle hybridization; $\Gamma \propto T$) [25,26]. Considering the nondegenerate character of the distribution (the absence of the Pauli blocking effect) it is to be expected that all free holes (with concentration $n_{2D} \sim 10^{14}$ cm$^{-2}$) take part both in conductivity and scattering processes [18]. In our case, the main carrier-scattering mechanism at an optimal doping is the electron–electron interaction, i.e., the scattering of holes on HL centres occupied by electrons. It is to be recalled that such a centre is an analogue of molecular ion $H_2^+$ (i.e., two electrons on the centre plus a hole coupled with this centre). The scattering frequency of a free hole on occupied HL centres is proportional to their concentration $N_{HL}^{oc}$ (which is equal to the concentration of free holes $n$) and the volume of the phase space for scattered holes $\Gamma \propto T$. That is,

$$\nu \propto N_{HL}^{oc} \Gamma.$$

The temperature dependence $\rho(T)$ in this model can be obtained from the Drude formula:

$$\rho = m^* \nu / n \cdot e^2,$$

(where $m^*$ is the effective mass). Inasmuch as $N_{HL}^{oc} = n$, we have $\rho \propto \Gamma \propto T$.

We should emphasize that the dependence (2) concerns the concentration of additional hole carriers generated at the expense of the occupation of HL centres additionally to localized doped carriers. In the field of underdoping, the contribution of additional hole carriers to the total concentration rapidly decreases with doping going down, in accordance with the decrease of the concentration of HL centres, and in heavily underdoped specimens becomes negligibly small. In this latter case, the total concentration of carriers ceases to depend on temperature [27]. In transition to the overdoping region, where small clusters and separate HL centres are "immersed" into normal metal, the contribution of additional carriers also drops down with doping owing to a decrease of the concentration of HL centres.

Let us consider now how the real electronic structure of cuprates changes due to transition from an insulator to a conductor. The initial structure of cuprates in the CT insulator phase is given in Fig. 1c. As the motion of an electron can be considered to be quasi-one-dimensional, the gaps and dispersions in the directions $(\pi,\pi)$, $(\pi,0)$ and $(0,\pi)$ are determined by the overlap integrals $t$ in the former case and $t'$ in two latter cases.

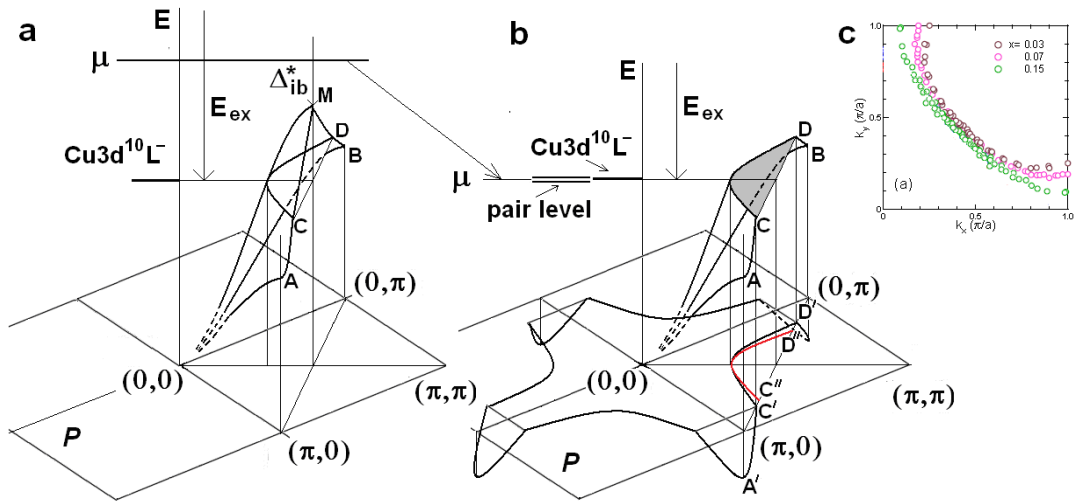

**Figure 4.** (**a**) The band structure of cuprates' CT phase. Electrons from the upper part of oxygen L band occupy exciton states $Cu3d^{10}L^-$. $CD$, an isoenergetic contour emerging at sectioning the dispersion curves by the dispersionless $Cu3d^{10}L^-$ level. To simplify the figure, the real remnant Fermi contour shown in Fig. 1d is straightened and coincides with the diagonal that joins points $(\pi,0)$–$(0,\pi)$. (**b**) In the CT phase two electrons of $Cu3d^{10}L^-$ on the pair level of an HL centre and two holes form a bound state. HL centres play the role of acceptors whose ionization leads to the emergence of free holes on the $CD$ isoenergetic contour and to a jump of the chemical potential. The $CD$ contour plays the role of a Fermi surface. $C'D'$, the projection of the $CD$ contour to the $(x,y)$ plane. Part of the dispersion curve CA lies lower than the $CD$ contour (correspondingly, its projection $C'A'$ is lower than $C'D'$ and forms a "large" pseudogap. As the doping is decreased, the dispersion along the CuO bonds goes down, which leads to an increase of the "large" pseudogap and a displacement of point $C$ to $C''$. Herewith, the positions of the centres of arcs $CD$ (and $C'D'$) determined by the values of dispersion in the diagonal direction remain invariable. (**c**) The Fermi surface in $La_{2-x}Sr_xCuO_4$ at an optimal doping ($x = 0.15$) and underdoping ($x = 0.07$ and $0.03$) from [28].

As we noted above, the main role of doping is to reduce locally the gap $\Delta_{ct}$ to a value of $\Delta^{*}_{ct} < E_{ex}$ for part of Cu sites. This is accompanied by a decrease of the average height of the barriers and, correspondingly, by an increase of dispersion in the direction $(\pi,0)$ and $(0,\pi)$ (Fig. 4a). With the electronic level of $Cu3d^{10}L^{-}$ higher than point $A$, the isoenergetic contour will have the form of arc $CD$ (or $C'D'$ in the projection to the plane) not reaching the boundaries of the 2D Brillouin zone (Fig. 1d). At $T > 0$, hole carriers emerge on these isoenergetic arcs, imitating the formation of a Fermi surface. Herewith, part of the dispersion curve $CA$ will lie energetically lower, forming a "large" pseudogap. Thus, the "large" pseudogap is a remnant of the interband gap on part of the Fermi contour around the anti-node directions. The "large" pseudogap exists both at $T > T_c$ and $T < T_c$ [29]. As the level of doping is decreased, the gap $\Delta^{*}_{ct}$ for part of Cu sites rises up to the previous value $\Delta_{ct}$ conforming to the undoped case. This will lead to a decrease of dispersion in the direction of Cu–O bonds at the dispersion preserved in the diagonal direction, where the width of the band is determined by only $t$. As the result, at a decrease of doping, the centre of the arc remains in the same position and its ends $C'$ and $D'$ are displaced in the direction of the diagonal towards points $C''$ and $D''$. This behaviour of the Fermi surface is consistent with the experiment [26] (Fig. 4c). Herewith, the value of the pseudogap determined by the distance from point $A$ to the plane of the Fermi contour will rise.

If in the cases shown in Fig. 2 the dopant concentrations are increased until the neighbouring trion complexes begin to overlap, this will lead to the delocalization of doped carriers and, correspondingly, to the transition to the overdoped (metal) phase. In the case of Fig. 2c, the transition to the overdoped phase with the doping level increasing will occur when the calculated charge $q^*$ per apical cation will significantly exceed the required value $|q^*| \sim |e|/4$. The latter takes place, in particular, in the HTSC $Ba_2Sr_2CaCu_2O_{8+x}$ ($0 < x < 2$), where a higher doping level as compared with $YBa_2Cu_3O_{6+x}$ ($0 < x < 1$) can be achieved. The superconductivity in the overdoped phase is realized only due to the preservation of Josephson-coupled islands of the CT phase in it. The coexistence of regions of the CT phase and "overdoped metal" can be the cause of a peak-dip-hump structure in the ARPES spectra of $Ba_2Sr_2CaCu_2O_{8+x}$ [30].

## 3. The mechanism of superconducting pairing. Fermi arcs, "small" pseudogap and pseudogap anomalies

According to the above said, incoherent transport in the CT phase of cuprates at $T > 0$ is performed by hole carriers emerging at the occupation of HL centres. However, at $T = 0$ the pair level is empty, and incoherent transport is impossible owing to the absence of free carriers. At the same time, such a system where an electron and a hole are present in each CT plaquette enables **coherent transport**, when all carriers of the same sign move coherently, as a single whole, e.g., a superconducting condensate. The latter is possible given a superconducting pairing. We believe that in the proposed pattern based on the formation of local CT excitons and HL centres such an interelectron attraction in the CT phase emerges due to the formation of a bound state of two electrons ($k\uparrow,-k\downarrow$), getting on the central Cu cations of an **unoccupied** HL centre, and two singlet holes, **with necessity** emerging herewith on the surrounding oxygen ions. Note that in this case the superconducting transition should be accompanied by a change of carrier type – from holes (at the incoherent transport in the normal state) to electrons (at the coherent transport in the superconducting state). This change of carrier sign in the superconducting transition is indeed observed in the experiment [31,32]. In addition, it should be also noted that, as each state in $L_d$ band (Fig. 3) is a superposition of band and exciton states, all electrons will be involved in pairing.

The proposed mechanism of superconducting pairing will also lead to the emergence of some other unusual properties in the system. First of all, consider the type of symmetry of the order parameter $\Delta(\boldsymbol{k})$, which is determined by the lattice symmetry and the pairing mechanism. As the movement of electrons in this system can be considered as quasi-one-dimensional, the superconducting order parameter (or pairing potential) $\Delta(\boldsymbol{k})$ will be determined by the frequency of transitions of electron pairs ($k\uparrow,-k\downarrow$) to HL centres, i.e., to copper sites. Herewith, $\Delta(k)$ will reach a maximum in the anti-node direction (i.e., along Cu–O bonds) and will turn to zero in the direction of O–O bonds. For an arbitrary direction of $k$ the frequency of electron transitions to Cu sites is proportional to the length of segment $AB$ in Fig. 5a. It is easy to see that $AB = \cos(k_x a) - \cos(k_y a)$, i.e., the order parameter will have a $d$-wave symmetry.

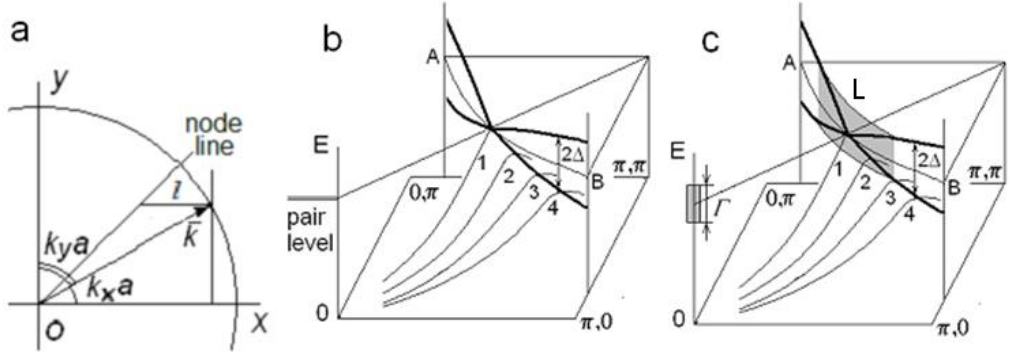

**Figure 5.** (**a**) Towards the determination of the order parameter in Cu-HTSCs. The frequency of transitions of an electron with momentum $k$ through Cu sites is proportional to the length of segment $l = \cos(k_x a) - \cos(k_y a)$ for a unit circle. (**b, c**) Formation of Fermi arcs as the result of the interaction of band electrons with HL centres. (b) $T = 0$. Lines 1–4, curves of the Bogoliubov quasi-particle dispersion. The upper branches are not shown. $AB$, a Fermi contour. The thick lines, a $d$-wave superconducting gap $\Delta(k)$. (c) $T > 0$. $\Gamma$, the width of a pair level of HL centres (the value of pair hybridization). The shaded belt (of length $L$), the region of depairing that occurs as the result of the transitions of pairs of electrons ($k_1, k_2$) to HL centres.

This type of symmetry, as we will see, also determines the possibility of depairing processes around node directions, which leads to the formation of Fermi arcs and a "small" pseudogap. Let us consider this mechanism in greater detail. Figure 5b shows an electron spectrum of cuprates at $T = 0$, which follows from the proposed model. Digits 1–4 designate some lower branches of Bogoliubov dispersion; $AB$, a Fermi contour on which a gap with a $d$-wave symmetry opens, which turns to zero in the direction ($\pi,\pi$). The "large" pseudogap in the vicinity of ($\pi,0$) and ($0,\pi$) is not shown.

A distinctive feature of the considered system is that, apart from heat excitations whose spectra are described by Bogoliubov dispersion curves (Fig. 5b), another two-particle excitations are also possible in the system. These pair excitations emerge owing to the pair hybridization of the level of HL centres with band states as the result of transitions of electron pairs ($k_1, k_2$) to the centres and back [25,26]. The value of pair hybridization $\Gamma$ (or inverse lifetime of a pair state, $\hbar/\tau$) depends on temperature [25,26].

$$\hbar/\tau \sim \Gamma \approx kT \cdot (V/E_F)^2 \qquad (3)$$

(here $V \sim 0.5$–1 eV is the one-particle hybridization constant; $E_F \sim 0.3$ eV, Fermi energy; $T$, temperature). Hence, $\Gamma \sim (5$–$10)kT$.

With temperature increased, the region of states from where real transitions of pairs of electrons to HL centres and back are possible (i.e., the region of states for which $2\Delta(k) < \Gamma$), will be extended from point ($\pi/2; \pi/2$) in the direction of the crest of dispersion, forming a belt of height $\Gamma$ and length $L$ along the Fermi contour $AB$ (Fig. 5c). For states in the band of pair hybridization, for which $2\Delta(k) < \Gamma$, transitions of pairs of electrons ($k_1, k_2$) to HL centres and back are accompanied by depairing and lead to the generation of Bogoliubov quasiparticles [33] within the limits of the belt of length $L$. These processes should have suppressed the superconducting order parameter at length $L$; however, due to the preservation of coherence in the system as a whole, a $d$-wave order parameter distinct from zero is preserved on the entire Fermi contour.

At the same time, with temperature increasing, the occupation of HL centres with real electrons rises to decrease the time-mean number of HL centres accessible for virtual transitions of electron pairs. With temperature increasing, when the population of HL centres reaches a critical value $\eta_c$, the phase coherence is disturbed, and the transition to the normal state takes place.

Herewith, the gap along arc *L* around the node direction (Fig. 5c) collapses as the result of the depairing processes. The length of arc *L* will increase proportionally to temperature [34]. At the same time, on the other parts of arc *AB* the "small" pseudogap, which now corresponds to incoherent pairing, is preserved. In the vicinity of points (π,0) and (0,π) (especially in the underdoped phase), the "large" pseudogap is preserved, *t*. Exactly the same pattern is observed in the ARPES [35].

Thus, as we noted above, the population of HL centres with real electrons rises with temperature (as does the concentration of additional free carriers). On the other hand, to implement the proposed pairing mechanism, HL centres should be **empty**, i.e., the occupation of HL centres as the result of pair hybridization is harmful for this mechanism of superconductivity. Therefore, the population of HL centres (and, correspondingly, temperature) should be sufficiently low to provide for the coherence to be established within the entire bulk. The superconductivity of a particular domain can be considered to require percolation over unoccupied HL centres, and temperature $T = T_c$ to correspond to a temperature at which unoccupied HL centres form an instantaneous percolation cluster.

This feature is especially pronounced in transition to the underdoped phase, when the percolation cluster of HL centres splits into finite clusters whose mean size decreases with doping going down [36]. Under these conditions, of increased significance is the effect of fluctuations of HL centres' population, which lead to the emergence of pseudogap anomalies. These anomalies emerge as a consequence of the fact that small clusters combining several HL centres fluctuate within the interval of $T_c < T < T^*$ ($T^*$, the pseudogap opening temperature) between the coherent superconducting and incoherent (normal) states. The number of clusters of HL centres occurring simultaneously in superconducting state, as the lifetime of this state, rise with temperature decreasing. Herewith, the experimentally measured $T_c$ has the meaning of temperature at which the percolation cluster of Josephson-coupled superconducting clusters of HL centres emerges. It is evident, however, that within some interval of temperatures $T_c < T < T^\nu$ there will be sufficiently long-lived and large (but not percolation) coherent superconducting clusters, where at $T > T_c$ the Nernst effect and giant diamagnetism can be observed [6,7]. As it follows from the above, the possibility of observing these anomalies is not determined immediately by the pseudogap, but is the result of the existence of fluctuationally occurring coherent superconducting regions. A significant feature of these effects associated with the existence of such short-lived superconducting clusters should be their dependence on the frequency of measurements [37].

## Conclusion

Thus, the given consideration enables describing the main properties of HTSC cuprates within the range from the CT insulator to normal metal. We sum up the main results.

**1. Doping mechanism.** The doped carrier self-localizes due to the formation of a trion complex, which comprises the carrier proper and CT excitons generated in its vicinity. The atomic complex of $CuO_4$ in the $CuO_2$ plane, where an exciton is generated, we called a CT plaquette. The arrangement of trion complexes in the basal plane of the crystal is determined by its structure and dopant position; therefore, the dopant concentration range corresponding to the existence of a percolation cluster of CT plaquettes (the CT phase) can be readily determined for each particular compound. It has been shown that for HTSC cuprates the found concentration ranges coincide exactly with the positions of superconducting domes on the experimental phase diagrams. That is to say, it is the CT phase that is the HTSC phase. Thus, the main result of doping is the transformation of the electronic structure of the $CuO_2$ plane from the state of a CT insulator to the CT phase, in which the band states can effectively interact with the exciton states.

**2. HL centres and Fermi surface.** Each pair of neighbouring CT plaquettes forms an HL centre, on which two electrons on Cu ions and two surrounding holes can form a bound state due to the possibility of singlet holes (occurring with necessity in the CT phase on the surrounding oxygen ions) to be in the space between Cu ions and be attracted simultaneously to two electrons present on these ions. The section of the band dispersion curves by the dispersionless pair level of HL centres produces an isoenergetic contour, on which hole carriers emerge as the result of transitions of part of electrons to partial ionization of HL centres (i.e., HL centres play the role of acceptors). It is this contour that plays the role of a Fermi surface, in essence not being such a surface. A consequence of such an unusual nature of the Fermi surface is the emergence of a "large" pseudogap in the vicinity of points $(\pi,0)$ and $(0,\pi)$, whose values increase with the doping decreasing.

**3. Mechanism of superconducting pairing.** The concentration of free holes depends on temperature, tending to zero at $T \to 0$. Nevertheless, in such a system, where there are an electron and a hole in each CT plaquette, a **coherent transport** is possible, when **all** carriers of the same sign move coherently, as a single whole or a condensate. Such a situation is possible given a superconducting pairing. We believe that the required interelectron attraction emerges due to the formation of a bound state of two electrons ($k\uparrow,-k\downarrow$) getting on the central Cu cations of an unoccupied HL centre and two singlet holes, **with necessity** present in the CT phase on the surrounding oxygen ions.

**4. Fermi arcs and the "small" pseudogap.** The emergence of a pair level of HL centres on the Fermi level leads to a two-particle hybridization of the band states with pair states of the centres, the value of which is $\Gamma \propto T$. As a consequence of $d$-symmetry, this hybridization at $T > 0$ is accompanied with depairing in the region of $k$ in the vicinity of the node directions, for which $\Gamma > 2\Delta(k)$. At some temperature, when the population of HL centres reaches a critical value, the phase coherence gets disturbed and the transition to the normal state takes place. Herewith, the gap along the arc, for which $\Gamma > 2\Delta(k)$, collapses as the result of depairing processes. At the same time, on the remaining part of the arc, $AB$, the gap is preserved (the "small" pseudogap), which now corresponds to the incoherent pairing.

**5. Pseudogap anomalies.** In the underdoped phase, the percolation cluster of CT plaquettes splits into finite clusters, whose average sizes decrease with the doping decreasing. Under these conditions, we observe a sharply increased effect of fluctuations in the population of HL centres, which lead to the emergence of pseudogap anomalies. These anomalies emerge as a consequence of the fact that small clusters combining several HL centres fluctuate within the interval $T_c < T < T^*$ between the coherent superconducting and normal states.

**Acknowledgements**

This work  was supported by Program  No 12 of the Presidium of the Russian Academy of Sciences "Fundamental problems of high-$T_c$ superconductivity".

**Author contributions**

Both authors contributed to the work presented in this paper. K.M. conceived the original idea and designed the model. O.I. provided critical feedback, analyzed of previously conducted experiments from literature and helped shape the research. Both K.M. and O.I. authors discussed the results and contributed to the final manuscript.

**Competing interests**

The authors declare no competing interests.